\documentclass[a4paper,10pt,preprint,sort&compress]{elsarticle}
\usepackage[english]{babel}
\usepackage{amsmath,amsthm}
\usepackage{amsfonts}
\usepackage{graphicx}
\usepackage{parskip}
\usepackage{color}
\usepackage{appendix}
\renewcommand{\appendixname}{Appendix}

\begin{document}

\title{A comparison of turbulence models in airship steady-state CFD simulations}
\author[setuh]{V. Voloshin\corref{cor1}}
\ead{v.voloshin@herts.ac.uk}
\author[setuh]{Y.K. Chen}
\ead{y.k.chen@herts.ac.uk}
\author[setno]{R. Calay}
\ead{rkc@hin.no}

\cortext[cor1]{Corresponding author: Tel: +44 (0)1707 284260; Fax: +44 (0)1707 285086 }

\address[setuh]{School of Engineering and Technology, University of Hertfordshire, College Lane, Hatfield, United Kingdom, AL10 9AB}
\address[setno]{Narvik University College, Lodve Langes gt. 2, Postboks 385, 8505 Narvik , Norway}

\begin{abstract}
The accuracy and resource consumption of the four different turbulence models based on the eddy viscosity assumption, namely, $k-\varepsilon$, two $k-\omega$ and Spallart-Allmaram models, in modeling airships are investigated. The test airship shape is a conventional shape. Three different angles of attack are considered. The results are checked against the wind tunnel experimental data. The resource consumption study is based on the benchmark of 1500 iterations. Based on all data obtained it is evident that Spallart-Allmaras model is the most optimal one in the majority of cases.
\end{abstract}


\maketitle

\section{Introduction}

Computational Fluid Dynamics allows scientists and engineers to test new airship shapes for their aerodynamic behaviour using only a computer. Airship simulations usually mean employing the flows with high Reynolds numbers. This fact requires using turbulence models to predict airship behaviour accurately.

The first interest in turbulence modeling appeared before the first computer was created, namely, in the beginning of last century. One of the pioneers was Prandtl with his mixing-length hypothesis published in 1925 \cite{Prandtl_en}. It was far from the modern models, but because all calculations were carried out by hands, the main focus was to reduce the number of operations as many as possible.

The first computers became available for the purposes of scientific research right after the Second World War. A new interest in turbulence modeling appeared at the same period of time because of the development of jet engines, supersonic aircrafts and some other technologies which required more accurate simulations. Many different turbulence models were developed during the period of 1940s-1960s. These were the first attempts of accurate prediction of near-wall layer turbulence flows.

But it was the beginning of 1970s when the modern turbulence models started to appear. The main achievement was the development of the parent 3 equation model by Hanjalic and Launder \cite{Hanjalic_Launder} and then the original 2 equation $k-\varepsilon$ model by Launder and Spalding \cite{Launder_Spalding}. The limitations of the latter model were soon found, in particular, inaccurate prediction of low Reynolds near-wall flows. The first modification of $k-\varepsilon$ model for a specific type of flow \cite{Jones_Launder} appeared in 1972, way before the paper on the finalised original model \cite{Launder_Spalding} was published. Other turbulence models for the accurate prediction of the boundary layer behaviour were developed at the same time (e.g. Ng and Spalding model \cite{Ng_Spalding} for turbulent kinetic energy $k$ and turbulent length scale $l$), but $k-\varepsilon$ with its modification became one of the most widespread models in the CFD world.

In 1980 another iconic turbulence model appeared. It is based on the same Boussinesq Hypothesis (or eddy viscosity assumption) and employs the same turbulent kinetic energy. But instead of dissipation rate $\varepsilon$ specific dissipation rate $\omega$ is used. The model was first introduced by Wilcox \cite{Wilcox}. Later Menter developed a modificated model called Menter SST $k-\omega$ model \cite{Menter}, which is used along with the original model and lots of its other modifications.

Another widespread turbulence model called Spallart-Allmaras model was introduced by Spalart and Allmaras in 1992 \cite{Spalart_Allmaras}. This model directly employs one equation for turbulent viscosity to close the Reynolds stress tensor in RANS. This model was specifically developed for external aerodynamic flows and thus is suitable for modeling airships.

So, in the beginning of 1990s many models and their modifications were available to simulate turbulent flows. But the choice of the model for some particular applications is not an obvious decision and usually is subject to a separate study. That was the main reason for many papers on turbulence models comparison. The cases considered in such papers can be divided into two types: developing flows and fully developed flows. In the first case strictly transient simulations can be used to see the dynamic effects. In the second type the fully developed flow can be studied using either transient or steady-state simulations.

The developing (or unsteady) flows are more complicated and, thus, much more papers have been published in this subject. Among the possible applications the following studies have been carried out: flows around cube or square (in 2D case) \cite{Murakami_Mochida, Murakami, Murakami2} and square cylinder \cite{Bosch_Rodi}, simulations of air jets \cite{Sanders_Sarh} and simulation of a working blood pump \cite{Song_Wood}.

The problems related to fully developed flows include, for example, pipe flows \cite{Hrenya_Bolio} and backward facing step problem \cite{Speziale_Thangam}. The latter is considered as a benchmark for turbulence models. One  particular interesting application in terms of fully developed flows is flow over airships. Omari et.al. \cite{Omari_Schall} did the comparison between RANS $k-\varepsilon$, LES Smagorinsky and VMS-LES models. They used a large prolate spheroid shape (airship like shape) with the angle of attack at $20^{\circ}$.

Modern increasing interest in airships requires better understanding of their aerodynamics and behaviour in order to make them safer and more efficient. Such tasks are solved using CFD nowadays. But as it was indicated before, turbulence model choice is not a trivial task. Thus a series of special comparison studies for different cases is necessary to give a guidance for scientists and engineers in this choice.

This work is aiming to solve this problem and help with such a choice for the case of fully developed flows over the airships for small, medium and large angles of attack. In this paper the performance of four turbulence models which work with eddy viscosity assumption, namely, two layer realizable $k-\varepsilon$, Standard and SST Menter $k-\omega$ and Spallart-Allmaras turbulence models, are compared. The results are validated against the experimental data obtained using wind tunnel. Commercial CFD code Star-CCM+ 6.04 was used for all the simulations.

\section{Theory} \label{turb_mod_sec}

All four models used for comparison are based on the Reynolds-Averaged Navier-Stokes equations. Classic form of this equation is
\begin{equation}
\rho\bar{u}_j  \frac{\partial \bar{u}_i }{\partial x_j}
= \rho \bar{f}_i
+ \frac{\partial}{\partial x_j}
\left[ - \bar{p}\delta_{ij}
+ \mu \left( \frac{\partial \bar{u}_i}{\partial x_j} + \frac{\partial \bar{u}_j}{\partial x_i} \right)
- \rho \overline{u_i^\prime u_j^\prime} \right ], \label{RANS}
\end{equation}
where $x_i$ are coordinates, $\bar{u}_i$ are respective time-averaged velocity components, $u'_i$ are respective fluctuations of the velocity components, $\bar{p}$ is a time-averaged pressure, $\rho$ is a density, $\bar{f}_i$ are external forces components, $\mu$ is a dynamic viscosity.

The unclosed terms $- \rho \overline{u_i^\prime u_j^\prime}$ are called Reynolds stresses. They are caused by the fluctuations of the velocity field. Turbulence models are used to close these terms. All four models considered in this paper are based on the eddy viscosity assumption, which can be written as follows
\begin{equation}
-\overline{u_i^\prime u_j^\prime} = 2\nu_tS_{ij}-\frac{2}{3}k\delta_{ij}, \label{eddy visc}
\end{equation}

where $\overline{S_{ij}} = \frac{1}{2}\left( \frac {\partial \overline{u_i}}{\partial x_j} + \frac{\partial \overline{u_j}}{\partial x_i} \right)$ is a mean rate of strain tensor, $\nu_t = \mu_t / \rho$ is a turbulent eddy viscosity, $k = \frac12 \overline{u_i^{\prime 2}}$ and $\delta _{ij}$ is the Kronecker delta.

The following four models are used for the study.

\subsection{RANS Realizable Two Layer $k-\varepsilon$ model}

  The standard $k-\varepsilon$ model consists of two transport equation
  \begin{equation}
  \frac{\partial}{\partial t} (\rho k) + \frac{\partial}{\partial x_i} (\rho k u_i) = \frac{\partial}{\partial x_j} \left[ \left(\mu + \frac{\mu_t}{\sigma_k} \right) \frac{\partial k}{\partial x_j}\right] + P_k + P_b - \rho \epsilon - Y_M + S_k \label{ke kinetic energy}
  \end{equation}
  and
  \begin{eqnarray}
  \frac{\partial}{\partial t} (\rho \varepsilon) + \frac{\partial}{\partial x_i} (\rho \varepsilon u_i) &=& \frac{\partial}{\partial x_j} \left[\left(\mu + \frac{\mu_t}{\sigma_{\varepsilon}} \right) \frac{\partial \varepsilon}{\partial x_j} \right] + \nonumber \\
  C_{1 \varepsilon}\frac{\varepsilon}{k} \left( P_k + C_{3 \varepsilon} P_b \right) &-& C_{2 \varepsilon} \rho \frac{\varepsilon^2}{k} + S_{\varepsilon}. \label{ke dissipation rate}
  \end{eqnarray}
  for turbulent kinetic energy $k$ and dissipation rate $\varepsilon$ respectively. In the equations (\ref{ke kinetic energy}) and (\ref{ke dissipation rate}) $P_k$ - production of $k$ term, $P_b$ - effect of buoyancy term, $Y_m$ - contribution of dilatation fluctuation to the overall dissipation rate, $S_k$ and $S_{\varepsilon}$ - user-defined source terms. The constants presented in the equations (\ref{ke kinetic energy}) and (\ref{ke dissipation rate}) have the following values:
  \begin{equation*}
  C_{1 \varepsilon} = 1.44, \; C_{2 \varepsilon} = 1.92, \; C_{\mu} = 0.09, \; \sigma_k = 1.0, \; \sigma_{\varepsilon} = 1.3.
  \end{equation*}
  It is not completely clear how to calculate the coefficient $C_{3 \varepsilon}$. The following formula has been employed
  \begin{equation*}
  C_{3 \varepsilon} = \tanh \frac{|v_b|}{|u_b|},
  \end{equation*}
  where $v_b$ is the velocity component parallel to the gravitational vector and $u_b$ is the velocity component perpendicular to the gravitational vector.

  The turbulent viscosity is represented by the formula
  \begin{equation}
  \mu_t = \rho C_{\mu} \frac{k^2}{\epsilon}. \label{ke visc model}
  \end{equation}

  However, the model used in this paper is a combination of realizable modification of standard model with two-layer approach. The realizable modification was introduced by Shih \cite{Shih_etal}. In this modification there is a new transport equation for the turbulent dissipation rate $\varepsilon$
  \begin{eqnarray}
  \frac{\partial}{\partial t} (\rho \varepsilon) + \frac{\partial}{\partial x_i} (\rho \varepsilon u_i) &=& \frac{\partial}{\partial x_j} \left[\left(\mu + \frac{\mu_t}{\sigma_{\varepsilon}} \right) \frac{\partial \varepsilon}{\partial x_j} \right] + \nonumber\\
  \rho C_1 S_\varepsilon &-& \rho C_2 \frac{\varepsilon^2}{k+\sqrt{\nu \varepsilon}} + C_{1\varepsilon} \frac{\varepsilon}{k} C_{3 \varepsilon} G_b + S_e \label{ke real dissipation rate}
  \end{eqnarray}

  and the coefficient $C_{\mu}$ in the model for the turbulent viscosity becomes variable and now depends on the mean flow and turbulence properties.

  Two layer approach introduced by Rodi \cite{Rodi} divides the whole domain into two layers. Turbulent dissipation rate $\varepsilon$ and turbulent viscosity $\mu$ are represented as a function of the wall distance in the layer adjacent to the wall. The values of $\varepsilon$ are computed using transport equation far from wall and blended smoothly with the near-wall function. There is no division into layers for the transport equation for the turbulent kinetic energy $k$, i.e. it is solved in the entire domain.

\subsection{RANS Standard $k-\omega$ model}

  The model was first introduced by Wilcox \cite{Wilcox} in 1988. Later he revised this model, but there is no validation for complex flows for revised version of the model. So, the original model is still widely used and referred to as the standard $k-\omega$ model.

  This is a two equation model and the main alternative to $k-\varepsilon$ model. Turbulent kinetic energy $k$ and specific dissipation rate $\omega$ are computed from two transport equations and used for turbulent viscosity modeling using the formula
  \begin{equation}
  \nu _T  = {\rho k \over \omega }. \label{ko visc model}
  \end{equation}

  The equation for the kinetic energy $k$ is
  \begin{equation}
  {{\partial \rho k} \over {\partial t}} + {{\partial \rho u_j k} \over {\partial x_j }} = \tau _{ij} {{\partial u_i } \over {\partial x_j }} - \beta ^* \rho k\omega  + {\partial  \over {\partial x_j }}\left[ {\left( {\mu  + \sigma_k \frac{\rho k}{\omega} } \right){{\partial \omega} \over {\partial x_j }}} \right] \label{ko kinetic energy}
  \end{equation}
  while the equation for the specific dissipation rate $\omega$ looks as follows
  \begin{eqnarray}
  {{\partial \rho \omega } \over {\partial t}} + {{\partial \rho u_j \omega } \over {\partial x_j }} &=& {\gamma\omega  \over k}\tau _{ij} {{\partial u_i } \over {\partial x_j }} - \beta \rho \omega ^2  + \nonumber \\
  & & {\partial  \over {\partial x_j }}\left[ {\left( {\mu  + \sigma_\omega \frac{\rho k}{\omega} } \right){{\partial \omega } \over {\partial x_j }}} \right] + \frac{\rho \sigma_d}{\omega} {{\partial k} \over {\partial x_j }} {{\partial \omega} \over {\partial x_j }}, \label{ko diss rate}
  \end{eqnarray}
  where
  \begin{equation*}
  \tau_{ij} = \mu_t \left( \frac{\partial u_i}{\partial x_j} + \frac{\partial u_j}{\partial x_i} - \frac23 \frac{\partial u_k}{\partial x_k} \delta_{ij} \right) - \frac23 \rho k \delta_{ij},
  \end{equation*}
  \begin{equation*}
    \varepsilon  = \beta ^* \omega k
  \end{equation*}
  and constants presented in (\ref{ko kinetic energy})-(\ref{ko diss rate}) have the following values
  \begin{equation*}
    \alpha  = {{5} \over {9}} \;~ \beta  = {{3} \over {40}} \;~ \beta^*  = {9 \over {100}}\;~ \sigma_{\omega}  = {1 \over 2}\; ~ \sigma_k  = {1 \over 2}.
  \end{equation*}

  This model works very well in viscous boundary layer (even in complicated cases) without any modifications. This is a significant advantage for the application of this model in the airship simulations. It also has a disadvantage in terms of the freestream and inlet $\omega$ value sensitivity. This disadvantage mostly affects internal flows, but still it can give some negative effect in case of airship simulations as well.

\subsection{RANS SST Menter $k-\omega$ model}

  The main disadvantage of the standard $k-\omega$ model was addressed by Menter in \cite{Menter}, where he introduced his modification for the standard model which was supposed to solve the main problem of the model leaving all its advantages. The approach is that standard $k-\omega$ model is used in the near-wall layer while the far field is computed using $k-\epsilon$ model with a special function, which blends the result of the two models together. There are other minor modifications to the standard model in SST model, for example, the last term in the transport equation for the specific dissipation rate $\omega$ becomes $2(1-F_1) (\rho \sigma_{\omega 2}/\omega) {{\partial k} \over {\partial x_j }} {{\partial \omega} \over {\partial x_j }}$. Again, the details can be found in the original Menter's paper \cite{Menter}.

\subsection{RANS Standard Spallart-Allmaras model}

  The Spallart-Allmaras model uses a single transport equation for modified turbulent viscosity $\tilde{\nu}$
  \begin{eqnarray}
  \frac{\partial \tilde{\nu}}{\partial t} + \frac{\partial u_j \tilde{\nu}}{\partial x_j} &=& C_{b1} [1 - f_{t2}] \tilde{S} \tilde{\nu} + \frac{1}{\sigma} \{ \nabla \cdot [(\nu + \tilde{\nu}) \nabla \tilde{\nu}] + C_{b2} | \nabla \nu |^2 \} - \nonumber \\
  & & \left[C_{w1} f_w - \frac{C_{b1}}{\kappa^2} f_{t2}\right] \left( \frac{\tilde{\nu}}{d} \right)^2 + f_{t1} \Delta U^2. \label{sa eq}
  \end{eqnarray}

  The turbulent viscosity is modelled using the following formula
  \begin{equation}
  \nu_t = \tilde{\nu} f_{v1},
  \end{equation}
  where $f_{v1} = \chi^3/(\chi^3 + C^3_{v1})$ and $\chi := \tilde{\nu}/\nu$. The notation and constant values for equation (\ref{sa eq}) can be found in \ref{sa_notation}.

  This model was specifically developed for the aerospace industry, which gives some hopes for accurate and not so resource-intensive (than the other three models) solution. Nevertheless, the standard model is usually considered as Low Reynolds number flow model, even the authors of the original model have shown that it works well for attached flows and flows with mild separation, which is exactly the case of airship simulations. That is the main reason for choosing the standard model over modified high Reynolds number model.

\section{Experimental and simulation approach}

\subsection{Airship shape and environment parameters}

\begin{figure}[h]
  \includegraphics[scale=0.25]{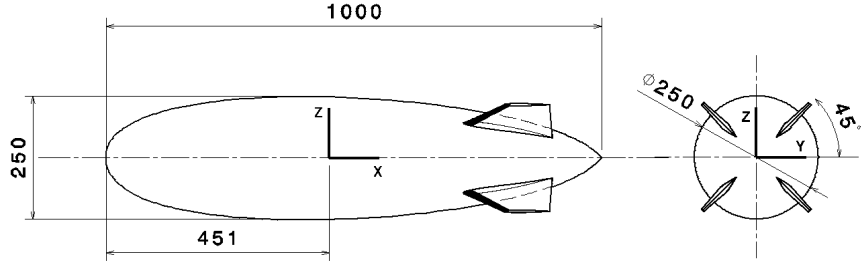}
  \caption{The model shape (dimensions are in mm)}.\label{shape}
\end{figure}

The shape of the model for all the simulations is a classic Zeppelin shape with 4 fins, all at a $45^\circ$ to the vertical and horizontal longitudinal planes (see Fig. \ref{shape}). The fineness ratio of the airship is 4:1 and the total chord length is 1 m. The centre of gravity is located on the axis of symmetry of the airship, the distance from the nose is 0.451 m. The simulations were conducted for models with for three angles of attack, namely $-0.4^\circ$, $11.62^\circ$ and $35.62^\circ$.

The ambient medium is a gas with the following parameters:
 \begin{itemize}
   \item constant density: $\rho=1.204 ~kg/m^3$,
   \item dynamic viscosity: $\mu = 1.789\cdot10^{-5} ~ Pa\cdot s$,
   \item ambient pressure at the sea level: $p_0 = 101325~Pa$,
   \item airship velocity: $v = 37 ~ m/s$.
 \end{itemize}

The coordinate system originates at the centre of gravity and consists of three axes X, Y and Z fixed to the body of the airship. The origin is in the centre of gravity. X represents the axis of symmetry of the airship looking backwards, Z is an axis perpendicular to X and looking up, Y makes up a right-handed coordinate system.

The air domain was chosen as a block and in cross section had the same sizes as the wind tunnel cross section in the test zone. The details about the wind tunnel tests see in Section \ref{exp_setup}. The boundary conditions again simulate the wind tunnel behaviour but with some modifications. The front surface of the air domain is chosen as the velocity inlet and the velocity was set to $v = 37~ m/s$. The back surface was simulating the outlet with ambient pressure $p_0 = 101325~Pa$. The sidewalls of the block were made as slip surfaces and the surface of the airship itself was made as non-slip one.

The results are presented as the aerodynamic coefficients rather than the forces and moments. For the coefficients, the following reference values were chosen:
 \begin{itemize}
   \item reference density: $\rho=1.204 ~kg/m^3$,
   \item reference area: $S = 0.101 ~m^2$,
   \item reference length: $l = 0.318 m$,
   \item reference velocity: $v = 37 ~ m/s$.
 \end{itemize}
All the simulations are of the steady-state type. We required each simulations to make 1500 iterations.
\subsection{Experimental setup} \label{exp_setup}

The experimental data was obtained using a closed wind tunnel with a 2x2 meters cross section in the test area. The airship sizes and the flow velocity was the same as for the simulation. The airship model was made out of polystyrene foam and fixed in supposed centre of gravity (see above). The sizes and the shape of the model are exactly the same as can be seen in the Figure \ref{shape}. All the experiments were duplicated and an average of two values was taken as final result.

\subsection{Mesh parameters}

A polyhedral mesh is a usual choice for the aerospace application. In particular, airships have quite steep filleted or rounded surfaces and polyhedral mesh is the best type of mesh for such shapes. Near-wall prism mesh consisting of 10 layers was used for more accurate simulation of the boundary layer flows. The air domain for each simulation contained approximately 4.5 million cells. The mesh is built in such a way that Y+ values in the nearest to the wall surface cells are around unity.

\section{Results and discussion}
\begin{figure}[h]
  \begin{picture}(450,250)
  \put(-20,-10){\includegraphics[scale=0.4]{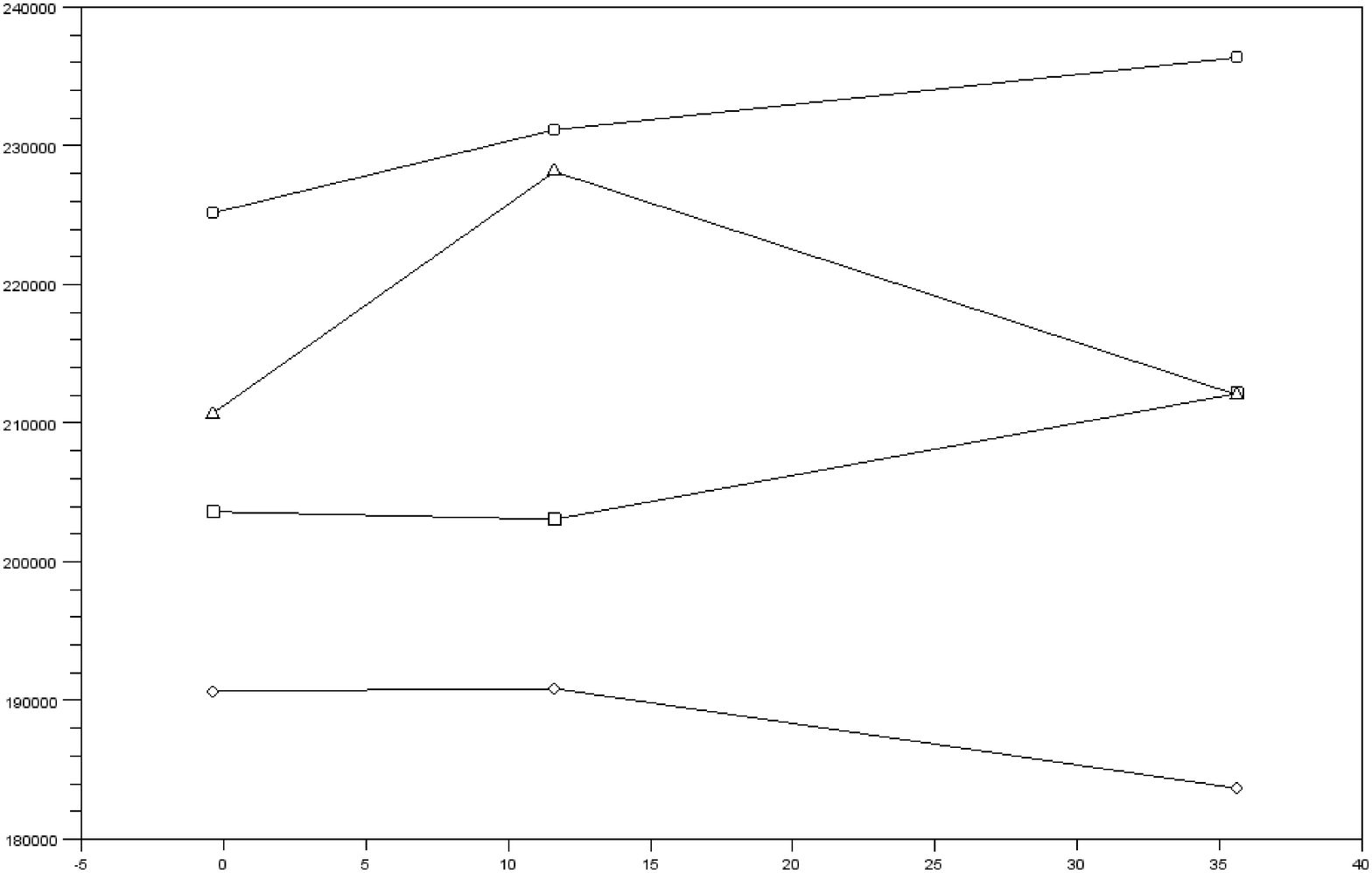}}
  \put(360,0){$\alpha,{}^{\circ}$ }
  \put(-10,240){\textit{cpu time, s}}
  \end{picture}
  \caption{CPU time required for doing a benchmark of 1500 iterations by Realizable two-layer $k-\epsilon$ ($\bigcirc$), Standard $k-\omega$ ($\square$), Menter SST $k-\omega$ ($\triangle$) and Spallart-Allmaras ($\lozenge$) models.}\label{cpu time pic}
\end{figure}

The results are presented in this section in Tables \ref{restab0040}-\ref{restab3562} for small, medium and large angles of attack. Each table contains two force coefficients $c_x$ and $c_z$ (the forces in the directions of the axes X and Z respectively) from wind tunnel test and CFD simulations using 4 different turbulence models described in section \ref{turb_mod_sec} together with the errors in comparison with experimental data. CPU time used for doing the benchmark of 1500 iterations can also be found in the tables as well as on the Figure \ref{cpu time pic}. Comparison between experiment results and modeling solutions using all 4 turbulence models are presented in a graphical form in Figures \ref{cx pic} and \ref{cz pic}.

\subsection{Angle of attack $\alpha = -0.4^{\circ}$} \label{sec0040}

\begin{table}[h]
\begin{tabular}{p{4cm}ccccc}
  \hline
  {} & $c_x$ & $c_z$ & $c_x$ error & $c_z$ error & CPU time, sec \\
 \hline
  Experimental results & 0.0419 & 0 & n/a & n/a & n/a\\
 \hline
  Realisable Two-layer $k-\varepsilon$ model & 0.039 & -0.008 & 6.9\% & n/a & 225157 \\
 \hline
  Standard $k-\omega$ model & 0.037 & -0.008 & 11.7\% & n/a & 203614 \\
 \hline
  Menter SST $k-\omega$ model & 0.035 & -0.008 & 10.0\% & n/a & 210695 \\
 \hline
  Standard Spallart-Allmaras model & 0.039 & -0.008 & 6.9\% & n/a & 190664 \\
  \hline
\end{tabular}
\caption{The result comparison table for the angle of attack $\alpha = -0.4^{\circ}$}\label{restab0040}
\end{table}

\begin{figure}[h]
  \begin{picture}(450,250)
  \put(-20,-10){\includegraphics[scale=0.4]{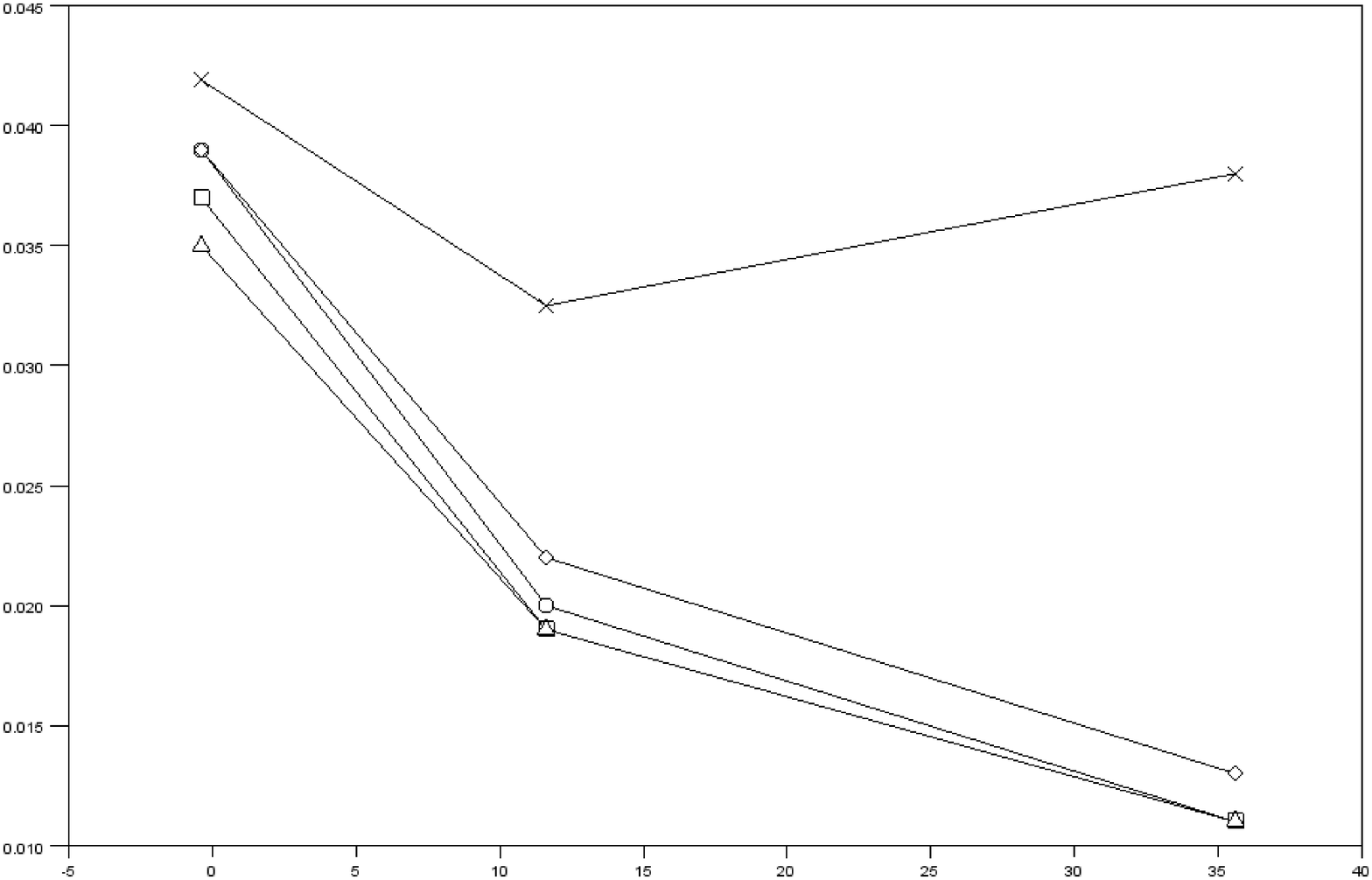}}
  \put(355,0){$\alpha,~{}^{\circ}$ }
  \put(-10,240){$c_x$}
  \end{picture}
  \caption{Comparison between the results for the coefficient $c_x$ given by the experiment ($\times$) and by Realizable two-layer $k-\epsilon$ ($\bigcirc$), Standard $k-\omega$ ($\square$), Menter SST $k-\omega$ ($\triangle$) and Spallart-Allmaras ($\lozenge$) turbulence models.}\label{cx pic}
\end{figure}

As can be seen from Table \ref{restab0040} the results for the $c_z$ are the same across all 4 models. The small deviations from the zero value can be easily explained by the errors caused by CAD importing and mesh generating procedures, which result in small asymmetry of the body. So, in the case of small angle of attack only the coefficient $c_x$ is compared. As shown in Figures \ref{cx pic}-\ref{cz pic} two models, namely $k-\varepsilon$ and Spallart-Allmaras turbulence models, provide the best results with the accuracy of 6.9\%. The other two models, which are actually standard and modified by Menter $k-\omega$ turbulence models, show much worse results with the accuracy of 11.7\% and 16.5\% respectively.
Nevertheless both $k-\varepsilon$ and Spallart-Allmaras turbulence models give very good results, the latter model required about 15\% less CPU time to finish 1500 iterations (see Figure \ref{cpu time pic}). It is can be easily explained by the fact that Spallart-Allmaras model is a one-equation model rather than two-equation $k-\varepsilon$ model.

It is suggested that in steady-state simulations of the airships flying with small (near-zero) angles of attack, the best turbulence model in terms of accuracy and required time and resources is Spallart-Allmaras model.

\subsection{Angle of attack $\alpha = 11.62^{\circ}$} \label{sec1162}

\begin{table}[h]
\begin{tabular}{p{4cm}ccccc}
  \hline
  {} & $c_x$ & $c_z$ & $c_x$ error & $c_z$ error & CPU time, sec \\
 \hline
  Experimental results & 0.0325 & 0.233 & n/a & n/a & n/a  \\
 \hline
  Realisable Two-layer $k-\varepsilon$ model & 0.020 & 0.248 & 38.5\% & 6.4\% & 231156 \\
 \hline
  Standard $k-\omega$ model & 0.019 & 0.249 & 41.5\% & 6.9\% & 203065 \\
 \hline
  Menter SST $k-\omega$ model & 0.019 & 0.249 & 41.5\% & 6.9\% & 228142 \\
 \hline
  Standard Spallart-Allmaras model & 0.022 & 0.247 & 32.3\% & 6.0\% & 190851 \\
  \hline
\end{tabular}
\caption{The result comparison table for the angle of attack $\alpha = 11.62^{\circ}$}\label{restab1162}
\end{table}

The case of the medium angle of attack gives a bit different picture. In Section \ref{sec0040} it was able to distinguish two best turbulence models in terms of accuracy based on comparing the coefficient $c_x$, in this case almost identical results (see Table \ref{restab1162}) for the coefficient $c_x$ are presented.

The results for the coefficient $c_z$ are expectedly less accurate. This can be easily explained by the fact that when the biggest contribution to the coefficient $c_z$ is made by pressure forces (about 98\%), the coefficient $c_x$ is mostly (about 90\%) made of shear forces. Pressure forces are much easier to predict than shear ones. Thus it is much more complicated to get the same level of accuracy for the coefficients significantly impacted by the viscous effects as for the ones containing mostly pressure effects.

\begin{figure}[h]
  \begin{picture}(450,293)
  \put(-20,-10){\includegraphics[scale=0.4]{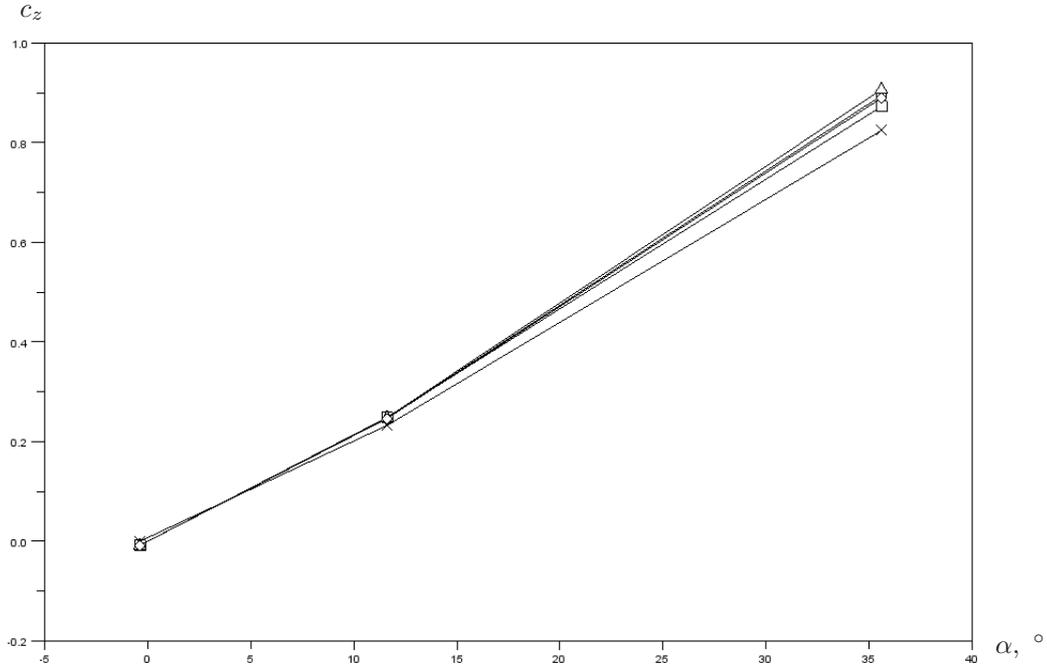}}
  \put(355,0){$\alpha,~{}^{\circ}$ }
  \put(-10,240){$c_z$}
  \end{picture}
  \caption{Comparison between the results for the coefficient $c_z$ given by the experiment ($\times$) and by Realizable two-layer $k-\epsilon$ ($\bigcirc$), Standard $k-\omega$ ($\square$), Menter SST $k-\omega$ ($\triangle$) and Spallart-Allmaras ($\lozenge$) turbulence models.}\label{cz pic}
\end{figure}

So, the result for medium angles of attack gives a very good opportunity to compare the accuracy of near-wall layer treatment by 4 different turbulence models with the same mesh conditions. Table \ref{restab1162} shows that the best results for the viscous effects are shown by the Spallart-Allmaras model. It is remarkable that the same model provides the best results for the pressure forces, or, in the other words, the highest accuracy level for the coefficient $c_z$. The result comparison can be seen in the Figures \ref{cx pic} and \ref{cz pic}.

Taking into account the fact, that the Spallart-Allmaras model uses the least CPU time (about 17\% less than $k-\varepsilon$ model, which is the next in accuracy), it is evident that Spallart-Allmaras turbulence model is the best model for the steady-state airship simulations with medium angles of attack.

\subsection{Angle of attack $\alpha = 35.62^{\circ}$} \label{sec3562}

\begin{table}[h]
\begin{tabular}{p{4cm}ccccc}
  \hline
  {} & $c_x$ & $c_z$ & $c_x$ error & $c_z$ error & CPU time \\
 \hline
  Experimental results & 0.038 & 0.825 & n/a & n/a & n/a\\
 \hline
  Realisable Two-layer $k-\varepsilon$ model & 0.018 & 0.895 & 52.6\% & 8.5\% & 236370 \\
 \hline
  Standard $k-\omega$ model & 0.011 & 0.873 & 71.1\% & 5.8\% & 212138 \\
 \hline
  Menter SST $k-\omega$ model & 0.011 & 0.907 & 71.1\% & 9.9\% & 212049 \\
 \hline
  Standard Spallart-Allmaras model & 0.013 & 0.890 & 65.8\% & 7.9\% & 183672 \\
  \hline
\end{tabular}
\caption{The result comparison table for the angle of attack $\alpha = 35.62^{\circ}$}\label{restab3562}
\end{table}

For the case of the large angles of attack it is clear (from Table \ref{restab3562}) that with the given mesh (especially, in the near-wall layer) all 4 models show quite poor accuracy for the coefficient $c_x$, which is contributed by viscous forces for 75\%. Thus, the performance in terms of viscous effects prediction is quite poor. In order to increase the accuracy of predicting viscous forces using the given turbulence models the mesh in the near-wall region must be refined.

But the pressure forces have much more significant effect on the overall aerodynamic behaviour of the airship, i.e. the coefficient $c_z$ have got 98\% contribution from the pressure forces. Moreover, as the accuracy of predicting the viscous forces is almost the same, we can compare the turbulence models based on the pressure forces prediction, or, in other words based on the accuracy of the results for the coefficient $c_z$.

Looking at Table \ref{restab3562} one can conclude that the best performance (5.8\% accuracy) was shown by the standard $k-\omega$ model. It is interesting that the modified $k-\omega$ models has the worst accuracy of 9.9\%. $k-\varepsilon$ model requires the biggest amount of resources to make the benchmark of 1500 iterations and gives not very good accuracy of 8.5\%. The Spallart-Allmaras model, which shows the best performance in the other two cases (see Sections \ref{sec0040} and \ref{sec1162}), again shows nice results in terms of resource consumption, but the accuracy is not so good in comparison with the standard $k-\omega$ model. The comparison between the experimental data and the numerical solutions can be seen in Figures \ref{cx pic} and \ref{cz pic}.

So, in case of the large angles of attack, $k-\omega$ turbulence model is the best in terms of the accuracy, but although Spallart-Allmaras model shows less accurate results it uses about 13\% less CPU time for it (for the comparison, see Figure \ref{cpu time pic}). In this case we can not recommend only one model. The choice of the turbulence model should be based on the requirements for the accuracy of the results and costs of the calculations. In other words, the choice of the model should depend on whether one is ready to trade some extra resources for about 2\% of extra accuracy, which does not necessarily play a significant role in preliminary airship simulation.

\section{Conclusions}

The results presented in Tables \ref{restab0040}--\ref{restab3562} allow to make a conclusion that among the considered turbulence models Spallart-Allmaras turbulence model is the most optimal in terms of accuracy (see Figures \ref{cx pic} and \ref{cz pic}) and resource consumption (see Figure \ref{cpu time pic}) for the simulations of airships flying at small (near zero) and medium (about $10^{\circ}$) angles of attack.

In case of large angles of attack the standard $k-\omega$ model performs more accurate than Spallart-Allmaras model (again, see Figures \ref{cx pic} and \ref{cz pic}) but it uses significantly more CPU time (see Figure \ref{cpu time pic}). So, Spallart-Allmaras turbulence model is suggested as the main choice in the simulation of airships flying at large (about $35^{\circ}$) angles of attack as well.

\section{Acknowledgements}

The authors would like to acknowledge financial support from EU-FP7-MAAT-Project. The authors are grateful to Francisco Campos for stimulating discussions.

\appendix
\appendixpage

\section{Notation for the Spallart-Allmaras model transport equation equation} \label{sa_notation}

The turbulent viscosity is expressed as
\begin{equation}
  \nu_t = \tilde{\nu} f_{v1},
\end{equation}
where $f_{v1} = \chi^3/(\chi^3 + C^3_{v1})$ and $\chi := \tilde{\nu}/\nu$. The transport equation is
\begin{eqnarray}
  \frac{\partial \tilde{\nu}}{\partial t} + \frac{\partial u_j \tilde{\nu}}{\partial x_j} &=& C_{b1} [1 - f_{t2}] \tilde{S} \tilde{\nu} + \frac{1}{\sigma} \{ \nabla \cdot [(\nu + \tilde{\nu}) \nabla \tilde{\nu}] + C_{b2} | \nabla \nu |^2 \} - \nonumber \\
  & & \left[C_{w1} f_w - \frac{C_{b1}}{\kappa^2} f_{t2}\right] \left( \frac{\tilde{\nu}}{d} \right)^2 + f_{t1} \Delta U^2, \label{sa eq2}
\end{eqnarray}
where
\[
\tilde{S} \equiv S + \frac{ \tilde{\nu} }{ \kappa^2 d^2 } f_{v2},~ f_{v2} = 1 - \frac{\chi}{1 + \chi f_{v1}},
\]
\[
S = \equiv \sqrt{2 \Omega_{ij} \Omega_{ij}},~ \Omega_{ij} \equiv \frac{1}{2} ( \frac{\partial u_i}{\partial x_j} - \frac{\partial u_j}{\partial x_i} ),
\]
\[
f_{t1} = C_{t1} g_t \exp\left( -C_{t2} \frac{\omega_t^2}{\Delta U^2} [ d^2 + g^2_t d^2_t] \right), ~f_{t2} = C_{t3} \exp(-C_{t4} \chi^2)
\]
\[
f_w = g \left[ \frac{ 1 + C_{w3}^6 }{ g^6 + C_{w3}^6 } \right]^{1/6},~ g = r + C_{w2}(r^6 - r), ~ r \equiv \frac{\tilde{\nu} }{ \tilde{S} \kappa^2 d^2 }
\]
and $d$ is a distance to the closest surface.

The constants presented in the equation (\ref{sa eq2}) have the following values
\[
\kappa = 0.41;~\sigma = 2/3;~C_{b1} = 0.1355;~C_{b2} = 0.622;~C_{w1} = C_{b1}/\kappa^2 + (1 + C_{b2})/\sigma ;
\]
\[
C_{t1} = 1 ;~C_{t2} = 2 ;~C_{t3} = 1.1 ;~C_{t4} = 2;~C_{v1} = 7.1;~C_{w2} = 0.3 ;~C_{w3} = 2
\]

Further details about the model, notation or the constants can be found in \cite{Spalart_Allmaras}.

\bibliographystyle{unsrt}
\bibliography{paper}

\end{document}